\documentclass[aps,prd,showpacs,floatfix,superscriptaddress,nofootinbib,10pt,twocolumn]{revtex4}
\usepackage{color,amsmath,amssymb,graphicx,latexsym,subfigure}
\usepackage{threeparttable,multirow,txfonts,makecell,tabularx}
\DeclareMathAlphabet{\mathcal}{OMS}{cmsy}{m}{n}

\begin{document}
	\title{Constraining cosmological phase transitions with the Parkes Pulsar Timing Array}

    \author{Xiao Xue}
    \affiliation{CAS Key Laboratory of Theoretical Physics, Institute of Theoretical Physics, Chinese Academy of Sciences, Beijing 100190, China}
    \affiliation{School of Physical Sciences, University of Chinese Academy of Sciences, Beijing 100049, China}
    \affiliation{II. Institute of Theoretical Physics, Universit\"at Hamburg, 22761 Hamburg, Germany}

	\author{Ligong Bian  }
	  \email{lgbycl@cqu.edu.cn}
   \affiliation{Department of Physics, Chongqing University, Chongqing 401331, China}
\affiliation{Chongqing Key Laboratory for Strongly Coupled Physics, Chongqing 401331, China}
    
    \author{Jing Shu}
    \email{ jshu@mail.itp.ac.cn}
    \affiliation{CAS Key Laboratory of Theoretical Physics, Insitute of Theoretical Physics, Chinese Academy of Sciences, Beijing 100190, China}
    \affiliation{School of Physical Sciences, University of Chinese Academy of Sciences, Beijing 100049, China}
    \affiliation{CAS Center for Excellence in Particle Physics, Beijing 100049, China}
    \affiliation{School of Fundamental Physics and Mathematical Sciences, Hangzhou Institute for Advanced Study, University of Chinese Academy of Sciences, Hangzhou 310024, China}
    \affiliation{Center for High Energy Physics, Peking University, Beijing 100871, China}
    \affiliation{International Center for Theoretical Physics Asia-Pacific, Beijing/Hangzhou, China}
    
    \author{Qiang Yuan}
    \email{yuanq@pmo.ac.cn}
    \affiliation{Key Laboratory of Dark Matter and Space Astronomy, Purple Mountain Observatory, Chinese Academy of Sciences, Nanjing 210023, China}
    \affiliation{School of Astronomy and Space Science, University of Science and Technology of China, Hefei 230026, China}
    \affiliation{Center for High Energy Physics, Peking University, Beijing 100871, China}
    
        \author{Xingjiang Zhu}
    \email{zhuxj@bnu.edu.cn}
    \affiliation{School of Physics and Astronomy, Monash University, Clayton, VIC 3800, Australia}
    \affiliation{OzGrav: The ARC Centre of Excellence for Gravitational Wave Discovery, Hawthorn, VIC 3122, Australia}
    \affiliation{Advanced Institute of Natural Sciences, Beijing Normal University at Zhuhai 519087, China}
    
    \author{N. D. Ramesh Bhat}
    \affiliation{International Centre for Radio Astronomy Research, Curtin University, Bentley, WA 6102, Australia}
    
    \author{Shi Dai}
    \affiliation{Western Sydney University, Locked Bag 1797, Penrith South DC, NSW 1797, Australia}
    
    \author{Yi Feng}
    \affiliation{University of Chinese Academy of Sciences, Beijing 100049, China}
    
    \author{Boris Goncharov}
    \affiliation{School of Physics and Astronomy, Monash University, Clayton, VIC 3800, Australia}
    \affiliation{OzGrav: The ARC Centre of Excellence for Gravitational Wave Discovery, Hawthorn, VIC 3122, Australia}
    
    \author{George Hobbs}
    \affiliation{CSIRO Astronomy and Space Science, P.O. Box 76, Epping, NSW 1710, Australia}

    \author{Eric Howard}
    \affiliation{CSIRO Astronomy and Space Science, P.O. Box 76, Epping, NSW 1710, Australia}
    \affiliation{Macquarie University, Department of Physics and Astronomy, Sydney, NSW, 2109, Australia}
    
    \author{Richard N. Manchester}
    \affiliation{CSIRO Astronomy and Space Science, P.O. Box 76, Epping, NSW 1710, Australia}

    \author{Christopher J. Russell}
    \affiliation{CSIRO Scientific Computing, Australian Technology Park, Locked Bag 9013, Alexandria, NSW 1435, Australia}

    \author{Daniel J. Reardon}
\affiliation{OzGrav: The ARC Centre of Excellence for Gravitational Wave Discovery, Hawthorn, VIC 3122, Australia}
\affiliation{Centre for Astrophysics and Supercomputing, Swinburne University of Technology, P.O. Box 218, Hawthorn, VIC 3122, Australia}

\author{R. M. Shannon}
\affiliation{OzGrav: The ARC Centre of Excellence for Gravitational Wave Discovery, Hawthorn, VIC 3122, Australia}
\affiliation{Centre for Astrophysics and Supercomputing, Swinburne University of Technology, P.O. Box 218, Hawthorn, VIC 3122, Australia}

\author{Ren\'ee Spiewak}
\affiliation{Jodrell Bank Centre for Astrophysics, University of Manchester, Manchester M13 9PL, UK}
\affiliation{Centre for Astrophysics and Supercomputing, Swinburne University of Technology, P.O. Box 218, Hawthorn, VIC 3122, Australia}

\author{Nithyanandan Thyagarajan}
\affiliation{CSIRO Astronomy and Space Science (CASS), P. O. Box 1130, Bentley, WA 6102, Australia}

\author{Jingbo Wang}
\affiliation{Xinjiang Astronomical Observatory, Chinese Academy of Sciences, 150 Science 1-Street, Urumqi, Xinjiang 830011, China}

	\begin{abstract}
		A cosmological first-order phase transition is expected to produce a stochastic gravitational wave background. If the phase transition temperature is on the MeV scale, the power spectrum of the induced stochastic gravitational waves peaks around nanohertz frequencies, and can thus be probed with high-precision pulsar timing observations. We search for such a stochastic gravitational wave background with the latest data set of the Parkes Pulsar Timing Array. We find no evidence for a Hellings-Downs spatial correlation as expected for a stochastic gravitational wave background.
		Therefore, we present constraints on first-order phase transition model parameters.
		Our analysis shows that pulsar timing is particularly sensitive to the low-temperature ($T \sim 1 - 100$~MeV) phase transition with a duration $(\beta/H_*)^{-1}\sim 10^{-2}-10^{-1}$ and therefore can be used to constrain the dark and QCD phase transitions. 
	\end{abstract}
	\maketitle
	\section{introduction}
	
	The stochastic gravitational wave background (SGWB) is an important target of gravitational-wave astronomy after the successful detection of gravitational waves from 50 compact binary mergers by the LIGO and Virgo detectors \cite{abbott2016observation,abbott2019gwtc,GWTC-2}. The SGWB may come from the superposition of numerous unresolvable binary coalescences, or from the cosmological inflation, cosmological phase transition, or topological defects expected to form during cosmological phase transitions, such as cosmic strings. Since the discovery of gravitational waves in 2015, the study of SGWB has gained increasingly broad interest because of its close connection with particle physics and cosmology. Different gravitational wave experiments, such as LIGO and the upcoming Laser Interferometer Space Antenna (LISA), aim at the detection of SGWB at different frequency bands \cite{Aasi:2013wya,Audley:2017drz,Caprini:2019egz}. 
	
	The pulsar timing array (PTA) experiments measure the arrival times of radio pulses from a number of highly stable millisecond pulsars in the Milky Way, which enables the detection of gravitational waves of nanohertz frequencies. The three major PTA experiments, the Parkes PTA (PPTA; \cite{manchester2013parkes}), the European PTA \cite{EPTA16}, and NANOGrav \cite{NANOGrav12.5data} have been monitoring the times of arrival (ToAs) of dozens of pulsars on weekly to monthly cadence for over 10 years, and provide unique probes of the SGWB in the nanohertz band \cite{Lentati:2015qwp,Caprini:2018mtu,Arzoumanian:2018saf,Garcia-Bellido:2017aan,Barack:2018yly,Caprini:2010xv,Breitbach:2018ddu}. 
		
	 Whereas the observation of the 125 GeV Higgs boson at the Large Hadron Collider indicates that the phase transition in the Standard Model of particle physics is cross-over~\cite{ DOnofrio:2014rug}, a first-order phase transition (FOPT) is predicted in many new physics models beyond the Standard Model, such as the real and complex extended standard models~\cite{ Espinosa:2011ax,Profumo:2014opa,Jiang:2015cwa}\footnote{See Ref.~\cite{Profumo:2007wc} for the early study on basic features of the Electroweak phase transition in the singlet model and its collider phenomenology. }, two-Higgs doublet models~\cite{Andersen:2017ika,Bernon:2017jgv, Dorsch:2013wja,Basler:2016obg}, next-to-minimal supersymmetric model~\cite{ Kozaczuk:2014kva,Huber:2015znp,Bian:2017wfv}, composite Higgs models~\cite{Espinosa:2011eu,Bian:2019kmg}, or the hidden sector~\cite{Schwaller:2015tja}. If the FOPT occurs at a relatively low temperature, e.g. around $T_*\sim $~MeV, a SGWB with a peak frequency $f\sim10^{-9}-10^{-8}$ Hz is expected~\cite{Caprini:2010xv,Bian:2020urb}, which falls into the sensitive region of PTA experiments~\cite{foster1990constructing,hellings1983upper}. In this work, we perform a search for the SGWB from the first-order phase transition (FOPT) of the early Universe using the latest PPTA data set. We show that the PPTA data can probe a considerable proportion of the parameter space of the FOPT. We ignore contribution from other sources to the SGWB; thus, constraints on FOPT derived in this work can be considered as being conservative.

	\section{Gravitational waves from FOPT}
	
	There are three main sources of gravitational waves produced during the FOPT: bubble collisions~\cite{ Caprini:2007xq,Huber:2008hg}, sound waves~\cite{ Hindmarsh:2017gnf,Hindmarsh:2013xza,Hindmarsh:2016lnk,Cutting:2019zws,hindmarsh2015numerical}, and MHD turbulence~\cite{Caprini:2009yp}. It is believed that sound waves are dominate in many phase transition models. In this work we only consider the sound wave contribution to the energy density of the SGWB from the FOPT, for the sake of conservativeness. Sound waves contribution to the SGWB over a characteristic frequency band $f$ can be expressed as \cite{Caprini:2015zlo,Hindmarsh:2017gnf,Ellis:2018mja,Ellis:2019oqb}
	\begin{align}
	\Omega_{sw}(f)h^2 &= 2.65\times 10^{-6}v_w
	\left(\frac{H_*}{\beta}\right)\left(\frac{\kappa\alpha}{1+\alpha}\right)^2\left(\frac{100}{g_*}\right)^{1/3}\nonumber\\
	&\times	\left[(f/f_{sw})^3\left(\frac{7}{4+3(f/f_{sw})^2}\right)^{7/2}\right]\nonumber\\
	&\times \min\left[1,~(8\pi)^{1/3}v_w\left(\frac{H_*}{\beta}\right)\left(\frac{3}{4}\frac{\kappa\alpha}{1+\alpha}\right)^{-1/2} \right],\label{edensity}
	\end{align}
	where $g_*$ is the effective number of relativistic degrees of freedom; $\kappa$ is the fraction of the latent heat of phase transition transferred into the kinetic energy of plasma~\cite{Espinosa:2010hh};  $v_w$ is the bubble wall velocity(we take $v_w= 1$ in this work); $H_*$ is the Hubble parameter at the phase transition temperature $T_*$; $\alpha$ is phase transition strength, which largely determine the amplitude of the GW spectrum; $\beta$ is the inverse duration of the phase transition (or the phase transition rate parameter), which together with $T_*$ set the peak frequency of the GW spectrum. $f_{sw}$ is the peak frequency of the sound wave defined as
	\begin{align}
	f_{sw} &\simeq \frac{1.9\times 10^{-5}~\mathrm{  Hz}}{v_w}\left(\frac{\beta}{H_*}\right)\left(\frac{T_*}{100\textrm{ GeV}}\right)\left(\frac{g_*}{100}\right)^{1/6}\;.\label{freq}
	\end{align}
	We fix the value of $g_*$ at different $T_*$: $g_* = 100$ when $T_* > 0.2$ GeV, $g_* = 10$ when $0.1$ MeV $<T_* < 0.2$ GeV, and $g_* = 2$ when $T_* < 0.1$ MeV \cite{husdal2016effective}. The set of parameters $\alpha$ and $\beta$ are defined at the nucleation temperature and entirely determine the gravitational wave spectrum. We note that, despite $g_*$ changing drastically at the critical temperatures, the accuracy of our model is not significantly affected due to the small power in Eqs.~(\ref{edensity}) and (\ref{freq}).

\section{Data analysis}
	
	The PTA experiments record pulse ToAs from highly stable millisecond pulsars. The SGWB can be detected by searching for specific statistical correlations among timing residuals of different pulsars \cite{allen1999detecting,maggiore2000gravitational}. The noise model we use is based on a detailed single pulsar analysis \cite{goncharov2020identifying}, and it is publicly available from Github repositories {\tt ppta\_dr2\_noise\_analysis}\footnote{https://github.com/bvgoncharov/ppta\_dr2\_noise\_analysis} and {\tt enterprise\_warp}\footnote{https://github.com/bvgoncharov/enterprise\_warp}. We use the {\tt TEMPO2} software \cite{hobbs2006tempo2,edwards2006tempo2} to calculate the timing residuals, and {\tt enterprise}\footnote{https://github.com/nanograv/enterprise}, {\tt PTMCMC}\footnote{https://github.com/jellis18/PTMCMCSampler}, {\tt enterprise\_extensions}\footnote{https://github.com/nanograv/enterprise\_extensions} for the Bayesian analysis. We also use the Jet Propulsion Laboratory solar-system ephemeris DE436, and the TT (BIPM18) reference timescale published by the International Bureau of Weights and Measures (BIPM).

    \subsubsection{Noise Model} 
    
    We briefly introduce our noise models here; we refer interested readers to Ref.~\cite{goncharov2020identifying} for more details. The noise components can be categorized as white noise and red noise. We use white noise parameters {\tt EFAC} and {\tt EQUAD} to account for additional white noise that is not included in ToA uncertainties for each signal processor/receiver system \cite{van2011placing,liu2011prospects,liu2012profile}. For some pulsars with multi-channel ToAs, white noise parameter {\tt ECORR} is also included to account for the pulse phase jitter noise \cite{arzoumanian2015nanograv}. For red noise, there are power-law spin noise, dispersion measure (DM) noise, chromatic red noise, and band noise \cite{van2013understanding,van2015low,caballero2016noise,lentati2016spin,you2007dispersion,keith2013measurement}. The exact composition of red noise terms differs from one pulsar to another. The formalism of the pulsar intrinsic red noises can be found in the {\tt Supplemental Material}. The ``exponential dip" events are included for 4 pulsars. For each pulsar, we apply a single pulsar noise analysis first. Then in the global Bayesian analysis, we fix the white noise parameters at their best-fit values from the single pulsar analyses, and vary the red noise and Bayes ephemeris parameters. The priors of the noise parameters can be found in Ref.~\cite{goncharov2020identifying}.
    
    Recently, the NANOGrav group found strong statistical support for a common power-law (CPL) spectrum process in their 12.5-year data set \cite{Arzoumanian:2020vkk}.
    Following the same analysis procedure/assumptions, such a CPL process was also identified in the PPTA data \cite{Boris21PPTAcrn}. However, it was pointed out that similar noise processes that are intrinsic to individual pulsars might lead to a false detection of a common noise.
    In both studies, evidence for the Hellings-Downs (HD) spatial correlation as expected from a SGWB is insignificant, and the nature of this CPL process is yet to be determined.
    We refer interested readers to Ref. \cite{Boris21PPTAcrn} for discussion on the possible common noise process. Following these two analyses, we introduce a common-spectrum process that is described by a power-law spectrum with the same amplitude and power-law index across different pulsars.
    Note that this work searches for SGWB from the FOPT process, which differs from Ref. \cite{Boris21PPTAcrn} both physically and technically.

    \subsubsection{Correlation search for gravitational waves} 
    
    The energy density of a SGWB, $\Omega_{gw} h^2$, is related with the one-sided power spectral density, $S_{gw}(f)$, as
	\begin{equation}
	S_{gw}(f) = \frac{1}{12\pi^2} \frac{1}{f^5} \frac{3H_{100}^2}{2\pi^2}\Omega_{gw}(f)h^2,\label{gwb}
	\end{equation}
	where $H_{100} = 100{~ \rm km ~ s^{-1}~  Mpc^{-1}}$ and $h \approx 0.7$. The statistical correlation between timing residuals $\delta t$ of pulsar $I$ and pulsar $J$ is
	\begin{equation}\label{corr}
	\langle \delta t_i^I ~\delta t_j^J \rangle = \int \mathrm{d}f~ S(f)~\Gamma^{IJ}(\theta_{IJ})~ \cos\left[2\pi f \left(t_i - t_j\right)\right],
	\end{equation}
	where $t_i$ and $t_j$ are ToAs, $\Gamma^{IJ}(\theta_{IJ})$ is the HD correlation, and $\theta_{IJ}$ is the angle between two pulsars (when $I=J$, $\theta_{IJ}=0$, $\Gamma^{IJ}=1$). Eq. (\ref{corr}) is valid under the assumption that the gravitational wave is isotropic and unpolarized, and only the Earth terms are considered. The time-frequency method is adopted, where a finite number of Fourier frequencies are chosen to approximate the power spectrum for an acceptable computation efficiency. For the gravitational wave spectrum, we also consider the case when the pulsar auto-correlation is subtracted. Namely, we use an off-diagonal HD correlation $\Tilde{\Gamma}^{IJ}(\theta_{IJ})$, and when $I=J$, we have $\theta_{IJ}=0$, $\Tilde{\Gamma}^{IJ}_0=0$. The correlation between a pulsar's own ToAs is referred to as pulsar auto-correlation. We use terms ``full HD'' and ``no-auto HD'' to differentiate these two cases.

	\subsubsection{SGWB spectrum} 
	
	We first consider that the power spectrum of the SGWB is a free function at 20 logarithmically spaced frequency modes between $10^{-9}$ and $10^{-8}$ Hz. Therefore the integral in Eq. (\ref{corr}) is calculated as a discrete summation of 20 elements \cite{van2015low}. We focus on the low-frequency range because PTAs are more sensitive in this range. We treat $\sqrt{\Omega_{gw}(f_i) h^2}$ at each $f_i$ as an independent parameter, for which we adopt a uniform prior between $10^{-10}$ and $1$. The red noise parameters and the Bayes ephemeris parameters are also treated as free parameters in the Bayesian analysis. The Markov Chain Monte Carlo (MCMC) method is employed to compute the posterior distribution of the parameters. The prior ranges of model parameters are summarized in Table~S1 of the {\tt Supplemental Material}.

	The FOPT model is roughly a broken power law with a peak frequency given by Eq.~(\ref{freq}). Similar to the free spectrum case, we also use 20 logarithmically spaced frequency modes between $10^{-9}$ and $10^{-8}$ Hz in Eq.~(\ref{corr}), but with $\Omega_{gw}(f_i)h^2$ as predicted by parameters $T_*$, $\alpha$, and $(\beta/H_*)^{-1}$. 
	
	The Savage-Dickey formula \cite{dickey1971weighted} is used to calculate the Bayes factor, $\rm{BF}$, for nested models; in our case we wish to calculate the BF between the signal hypothesis and the noise hypothesis:
	\begin{equation}
	    \mathrm{BF} \equiv \frac{\mathcal{Z}_1}{\mathcal{Z}_0} = \frac{\mathrm{P}(\varphi = \varphi_0)}{\mathrm{P}(\varphi = \varphi_0|\boldsymbol{\rm D})},
	\end{equation}
	where $\mathcal{Z}_1$ ($\mathcal{Z}_0$) is the model evidence for the signal (noise) hypothesis, $\boldsymbol{\rm D}$ and $\varphi$ denote the data and the signal parameters, respectively, and $\varphi_0$ represents the noise-only model that is nested in the signal model (i.e., the signal has zero amplitude). $\rm{P}(\varphi = \varphi_0|\boldsymbol{\rm D})$ and $\rm{P}(\varphi = \varphi_0)$ are the posterior and prior probability of noise-only hypothesis, respectively. A nested model requires signal parameters $\varphi$ to be independent of noise parameters $\vartheta$, which is true for our FOPT model.

	\section{Results}

	\begin{figure}[htp!]
	\includegraphics[width=0.48\textwidth]{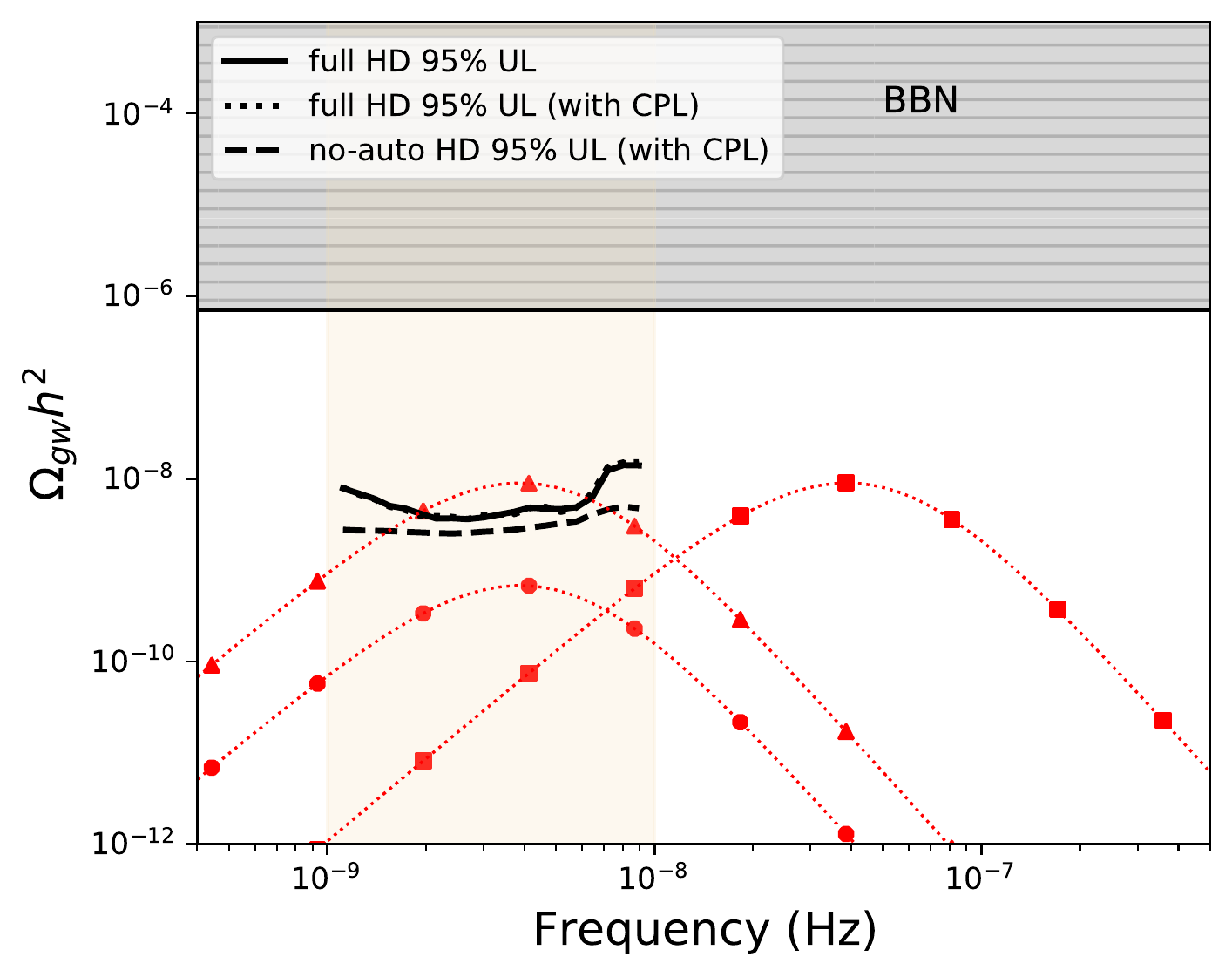}
	\caption{The 95\% credibility Bayesian upper limits on the energy density of a free-spectrum SGWB based on the PPTA data. The solid, dotted and dashed black curves indicate results under different assumptions (hypotheses H2, H3 and H4 in Table \ref{BayesFactors}, respectively, where the FOPT spectrum is substituted with the free spectrum). 
	The region excluded by BBN \cite{cyburt2005new} is shaded grey. Red curves with filled markers show examples of the spectrum predicted in the FOPT model, with $(\beta/H_*)^{-1}=0.1$, and varying parameters $(\alpha,\,T_*)=\{(0.2,\,3~\textrm{MeV})\ \textrm{(circle)},\,(0.5,3~\textrm{MeV})\ \textrm{(triangle)}, (0.5,\,30~\textrm{MeV})\ \textrm{(square)}\}$, respectively.}\label{free95}
	\end{figure}

    \begin{table*}[htp!]
    \caption{Description of hypotheses tested in this work and the Bayes factors between them.} \label{BayesFactors}
    \begin{tabular}{|l|c|c|c|c|c|c|}
    \hline
    \multirow{2}*{Hypothesis} & \multirow{2}*{\shortstack{Pulsar \\ noise}}& \multirow{2}*{\shortstack{CPL \\ process}} & \multirow{2}*{\shortstack{HD process\\ FOPT spectrum}} & \multirow{2}*{Bayes Factors} & \multicolumn{2}{c|}{Parameter Estimation (median and 1-$\sigma$ interval)}\\
    \cline{6-7}
    &&&&&$T_*$/MeV, $\alpha\times 10^{3}$, $\beta/H_*$&$A_{\rm CPL}$, $\gamma_{\rm CPL}$
    \\
    \hline
H0:Pulsar Noise   & \checkmark &  &  & &&\\
    \hline
    H1:CPL  & \checkmark & \checkmark &  & $10^{2.9}$ (against H0)&&$-14.45_{-0.64}^{+0.62}$,$3.31_{-1.53}^{+1.36}$\\
    \hline
    H2:FOPT  & \checkmark &  & \checkmark (full HD) & $10^{1.8}$ (against H0)&$7.4_{-4.7}^{+11.9}$, $271_{-92}^{+165}$, $9.9^{+11.4}_{-5.4}$&\\
    \hline
    H3:FOPT1  & \checkmark & \checkmark & \checkmark (full HD) & 1.04 (against H1)&$9.6_{-9.2}^{+232.2}$, $3.8_{-3.4}^{+27.9}$, $854^{+9622}_{-782}$&$-14.51_{-0.68}^{+0.64}$,$3.36_{-1.54}^{+1.39}$\\
    \hline
    H4:FOPT2 & \checkmark & \checkmark & \checkmark (no-auto HD) & 0.96 (against H1)&$10.9_{-10.6}^{+290.5}$, $3.2_{-2.8}^{+19.9}$, $1053^{+11256}_{-962}$&$-14.45_{-0.64}^{+0.62}$,$3.27_{-1.54}^{+1.37}$\\
    \hline
    \end{tabular}
    \end{table*}	
    
    We show in Fig.~\ref{free95} the 95\% credibility upper limits on the energy density of a SGWB in the $10^{-9}-10^{-8}$ Hz frequency range. The SGWB is characterised with a free spectrum. We note that the upper limits are almost unaffected by the inclusion of a CPL process.
    This is due to the choice of priors for the free-spectrum SGWB amplitude (uniform) and the CPL amplitude (log-uniform); see Table~S1 in the {\tt Supplemental Material} for details.
    We choose a uniform amplitude prior for the SGWB because it results in more conservative upper limits than the log-uniform prior.
    The upper limits are slightly more constraining for the case with only no-auto HD correlation included.
    Our results improve significantly upon the Big Bang nucleosynthesis (BBN \cite{cyburt2005new}) limit by two orders of magnitude in the frequency range considered in this work. 
    
    In the search for a FOPT signal, we consider three cases, listed as H2, H3 and H4 in Table \ref{BayesFactors}, respectively. In H2, we assume that there is no CPL process. In both H3 and H4, a CPL process is included. In H3, we include the full HD correlation induced by a SGWB, whereas in H4 only the off-diagonal correlation (denoted as no-auto HD) is used. For comparison, we also consider H1 where we assume there is only a CPL process.
    
    We summarize the hypotheses tested in this work and the Bayes factors between them in Table~\ref{BayesFactors}. First, being consistent with a separate analysis \cite{Boris21PPTAcrn}, we find a strong evidence ($\rm{BF}\sim10^{2.9}$) for a CPL process, if we follow the same analysis procedure/assumptions in Ref \cite{Arzoumanian:2020vkk}.
    We then search for evidence of phase-transition gravitational waves.
    Without adding a CPL term (H2), we obtain a BF of $10^{1.8}$ in favour of the FOPT signal, which is less significant than that of CPL process (H1). Nevertheless, we show the posterior distributions of the FOPT model parameters in Fig.~S1 of the {\tt Supplemental Material}.
    
    Since the origin of the apparent CPL process is still an open question, we take a conservative approach by assuming that there are unknown systematic errors that can be approximated by a CPL process\footnote{In the case that the apparent CPL process is due to a supermassive binary black hole (SMBBH) background, we find that the constraints on FOPT parameters are essentially unchanged; see the {\tt Supplemental Material}.}. Under hypotheses H3 and H4,
    the evidence of the FOPT signal is negligible ($\rm{BF}\sim1$).

\begin{figure*}[!t]
   	\includegraphics[width=1\textwidth]{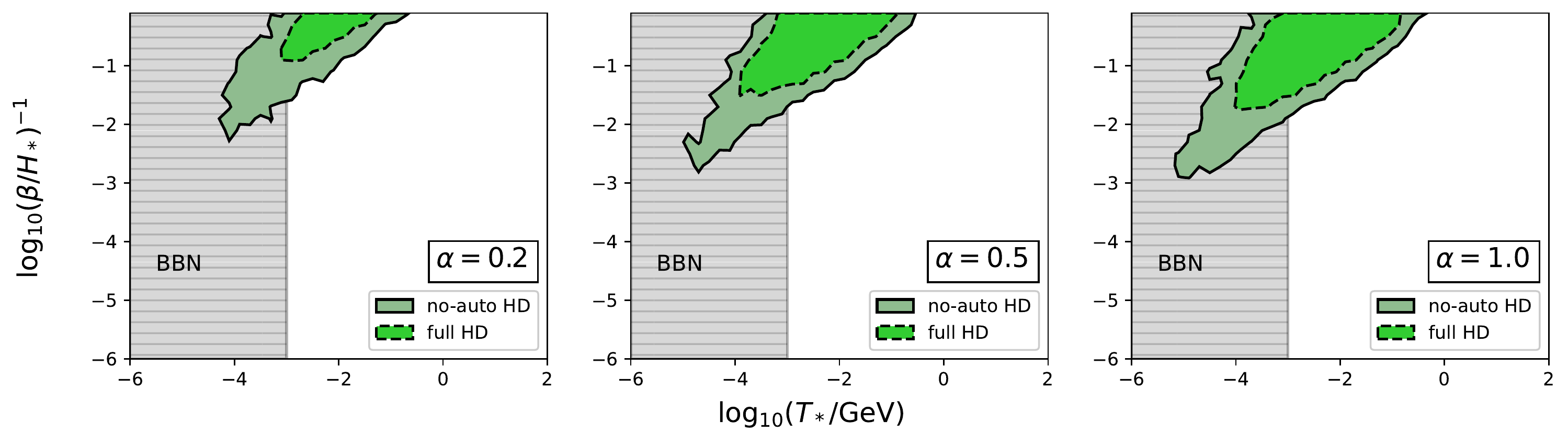}
	\caption{The PPTA exclusion contours (at 95\% C.L.) on FOPT parameters $\log_{10}(T_*/{\rm GeV})$ and $\log_{10}[(\beta/H_*)^{-1}]$, for $\alpha=0.2$, 0.5, and 1.0.}\label{FOPT95}
    \end{figure*}

	In Fig.~\ref{FOPT95}, we show the exclusion contours (at the 95\% credible level) on FOPT model parameters for $\alpha =$ 0.2, 0.5, and 1. There are a few noteworthy features.
	First, the constraints become tighter for a larger value of $\alpha$ (i.e., larger phase transition strength).
	Second, the excluded range of phase-transition temperature $T_*$ is greater for longer phase-transition duration (i.e., larger $(\beta/H_*)^{-1}$).
	Third, the excluded parameter space is larger for the case of no-auto HD correlation.
	This is because the FOPT models is slightly disfavoured by the data when auto-correlation of the SGWB is subtracted (see Table \ref{BayesFactors}).
	Compared with the BBN bound on the phase transition temperature ($T_*> 1 $ MeV), our results are more sensitive when $1\ {\rm MeV} <T_* < 100\ {\rm MeV}$.

	\section{Conclusion}
	Pulsar timing arrays provide a unique way to detect gravitational waves in the nanohertz frequency range ($10^{-9}$ Hz to $10^{-7}$ Hz), enabling the search for an SGWB potentially produced from cosmological phase transitions at low phase-transition temperatures ($\sim1$ MeV). In the PPTA second Data Release \cite{kerr2020parkes}, we search for the SGWB induced by first-order phase transitions of the early Universe. Assuming there is an unknown systematic error that can be approximated by a CPL process, we find no evidence for a Hellings-Downs spatial correlation induced by the phase-transition gravitational wave signal.
	We are able to exclude a significant parameter space of cosmological phase transitions that is not probed by previous observations. We find that phase transitions can be effectively constrained by the PPTA data for phase-transition temperatures of $\sim 1-100$~MeV and durations $(\beta/H_*)^{-1}\sim 10^{-2}-10^{-1}$, which are within predictions from low-scale dark phase-transition models \cite{Ratzinger:2020koh,Bian:2020urb,Breitbach:2018ddu} and cosmological first-order QCD phase transitions \cite{Caprini:2010xv}. The results therefore place constraints on the new physics admitting a first-order phase transition at low-energy scales in the early Universe, and provide an alternative way to probe particle physics at low-energy scales.
	
	\textit{Note added.}---Recently the NANOGrav collaboration also performed a search for the gravitational-wave background from the FOPT in their 12.5-year data set \cite{NANOGrav21fopt}. In that work, the gravitational-wave signal from phase transitions was proposed as a possible interpretation of the detected CPL process. As pointed out in Ref. \cite{Boris21PPTAcrn}, the detection of a CPL process might suffer from model misspecification. Therefore, we consider the feature found in our data to be a noise component of unknown origin and derive constraints on the FOPT parameters. Better understanding of the nature of the CPL process requires longer data sets and more pulsars which might be available through the International Pulsar Timing Array \cite{IPTAdr2} in the future.
 	
	\section{acknowledgements}
	We thank the referees for very useful comments.
	This work has been carried out by the Parkes Pulsar Timing Array, which is part of the International Pulsar Timing Array.
	The Parkes radio telescope (``Murriyang") is part of the Australia Telescope, which is funded by the Commonwealth Government for operation as a National Facility managed by CSIRO.
	This work is supported in part by the National Key Research and Development Program of China under Grant No. 2020YFC2201501.
	XX is supported by Deutsche Forschungsgemeinschaft under Germany's Excellence Strategy EXC2121 ``Quantum Universe'' - 390833306.
	LB is supported by the National Natural Science Foundation of China (NSFC) under the grants No. 12075041 and No. 12047564, the Fundamental Research Funds for the Central Universities of China (No. 2021CDJQY-011 and No. 2020CDJQY-Z003), and Chongqing Natural Science Foundation (Grant No. cstc2020jcyj-msxmX0814).
	JS is supported by the NSFC under Grants No. 12025507, No. 11690022, No. 11947302, by the Strategic Priority Research Program and Key Research Program of Frontier Science of the Chinese Academy of Sciences (CAS) under Grants No. XDB21010200, No. XDB23010000, No. XDPB15, No. ZDBS-LY-7003, and by the CAS Project for Young Scientists in Basic Research under Grant No. YSBR-006. 
	QY is supported by CAS and the Program for Innovative Talents and Entrepreneur in Jiangsu.
	XZ was supported by ARC CE170100004.
	SD is the recipient of an Australian Research Council Discovery Early Career Award, project DE210101738 funded by the Australian Government.
	RMS is the recipient of Australian Research Council Future Fellowship FT190100155. 

\bibliography{reference}
\bibliographystyle{ieeetr}

\clearpage

\begin{appendix}
\section*{Supplemental Material}

\subsection{Formalism of Pulsar Intrinsic Red Noise}\label{appendixnoise}
The spin noise (for all pulsar ToAs) is parameterized as a power-law of frequency
\begin{align}
    S_{\rm spin}(f) =  \frac{A_s^2}{12\pi^2} \left(\frac{f}{{\rm yr}^{-1}}\right)^{-\gamma_s}{\rm yr}^{3}.\label{spin}
\end{align}
The band noise is the same as spin noise, but only for ToAs of a certain photon frequency band,
\begin{align}
    S_{\rm band}(f) =  \frac{A_b^2}{12\pi^2} \left(\frac{f}{{\rm yr}^{-1}}\right)^{-\gamma_b}{\rm yr}^{3}.\label{band}
\end{align}
The dispersion measure (DM) noise applies to all pulsar ToAs, but correlates between different observing photon frequencies, as 
\begin{align}
    S^{ij}_{\rm DM}(f) = \frac{A_{\rm DM}^2}{12\pi^2} \left(\frac{f}{{\rm yr}^{-1}}\right)^{-\gamma_{\rm DM}}{\rm yr}^{3} \left(\frac{\nu_i}{\nu_{\rm ref}}\right)^{-2} \left(\frac{\nu_j}{\nu_{\rm ref}}\right)^{-2},\label{DM}
\end{align}
where $\nu_i$ and $\nu_j$ are photon frequencies of the $i$th and $j$th ToA, $\nu_{\rm ref}=1400$ MHz is a reference frequency. The chromatic red noise is 
\begin{align}
    S^{ij}_{\rm chrom}(f) = \frac{A_c^2}{12\pi^2} \left(\frac{f}{{\rm yr}^{-1}}\right)^{-\gamma_c}{\rm yr}^{3} \left(\frac{\nu_i}{\nu_{\rm ref}}\right)^{-n} \left(\frac{\nu_j}{\nu_{\rm ref}}\right)^{-n},\label{chrom}
\end{align}
where $n$ variates from pulsar to pulsar.

The CPL process for all pulsars is
\begin{align}
    S_{\rm CPL}(f) = \frac{A_{\rm CPL}^2}{12\pi^2} \left(\frac{f}{{\rm yr}^{-1}}\right)^{-\gamma_{\rm CPL}}{\rm yr}^{3}.\label{CPL}
\end{align}
The correlation $C_{ij}$ between ToA residuals is written as
\begin{align}
    C^{ij}_{\rm red} = \int_{f_L} {\tt d} f \left(\sum S^{ij}(f)\right) \cos\left(2\pi f (t_i - t_j)\right).
\end{align}
where $\sum$ represents the summation over all red contributions, including noises, CPL process and gravitational wave background signals. For pulsar noises  Eq.~(\ref{spin}-\ref{chrom}), $f_L = 1/T$, $T$ is the duration of observation for the pulsar, The time-frequency method is adopted to approximate the integrals, and 30 linearly spaced frequency modes are chosen, i.e., $f_i = i/T$, $i = 1...30$. For the CPL process Eq.~(\ref{CPL}) and all gravitational wave background signals, we use 20 logarithmically spaced frequency modes between $10^{-9}$ Hz and $10^{-8}$ Hz.

\begin{table*}[htb]
\centering
\caption{Description and priors of the model parameters. The model of pulsar intrinsic noise we use is the same as ref. \cite{goncharov2020identifying}, therefore no redundant description will be given here.}
\label{table-parameter}
\begin{tabularx}{\textwidth}{|X|X|X|X|}
\hline
& Parameter & Prior & Description \\
\hline
CPL Process Parameters
& $A_{\rm CPL}$ & log-U[$10^{-18}$,$10^{-11}$] & one per PTA\\
&$\gamma_{\rm CPL}$ & U[$0$,$7$] & one per PTA\\
\hline
\multirow{2}*{\shortstack{FOPT Parameters}}
&$\alpha$ & log-U[$10^{-5}$,$1$] & one per PTA\\
&$(\beta/H_*)^{-1}$ & log-U[$10^{-6}$,$1$] & one per PTA\\
& $T_*$ (GeV) & log-U[$10^{-6}$,$10^2$] & one per PTA\\
\hline

Free Spectrum Parameters &$\sqrt{\Omega_{gw}(f_i)h^2}$&U[$10^{-10}$,$10^{0}$]&20 per PTA\\
\hline
\multirow{6}*{\shortstack{{\tt BayesEphem}~[12] Parameters }}
&$z_{\rm drift}$ &U[$-10^{-9}$,$10^{-9}$]& one per PTA \\
&$\Delta M_{\rm Jupiter}$&N$(0,1.5\times10^{-11})$& one per PTA\\
&$\Delta M_{\rm Saturn}$&N$(0,8.2\times10^{-12})$& one per PTA\\
&$\Delta M_{\rm Uranus}$&N$(0,5.7\times10^{-11})$& one per PTA\\
&$\Delta M_{\rm Neptune}$&N$(0,7.9\times10^{-11})$& one per PTA\\
&${\rm PCA}_i$&U[$-0.05$,$0.05$]& six per PTA\\
\hline\hline
\end{tabularx}
\end{table*}

\begin{figure*}[!htb]
\includegraphics[width=1\textwidth]{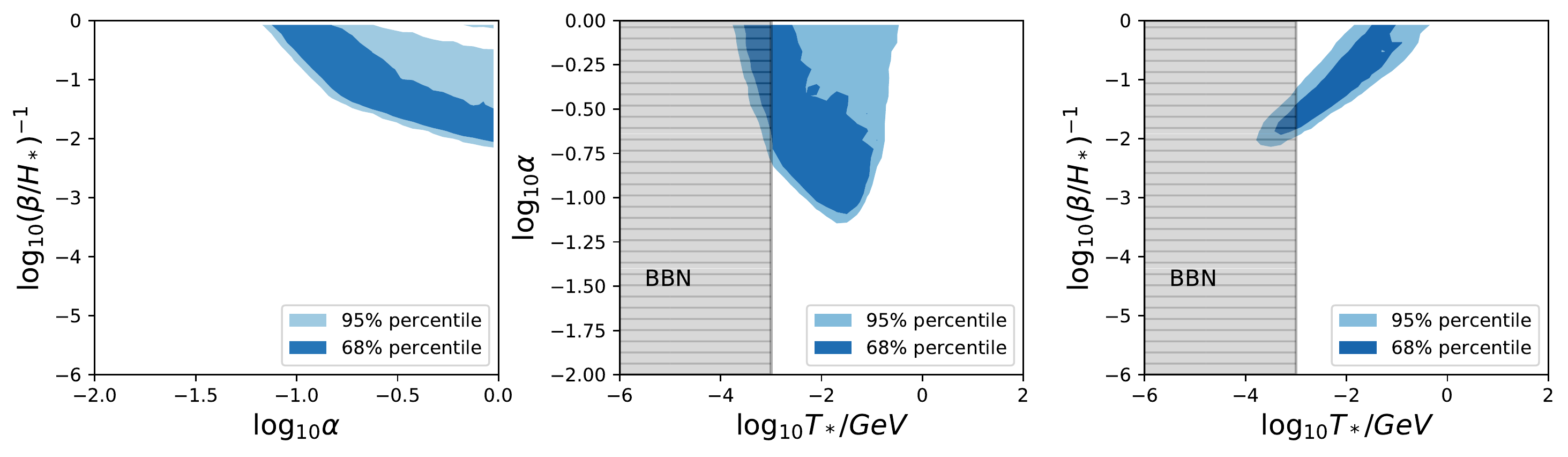}
\caption{Posterior probability distributions of the FOPT parameters assuming no CPL process (under the H2 hypothesis in Table I). For each panel the other parameter is marginalized within its prior range. The excluded region from BBN \cite{cyburt2005new} is shown as a grey shaded band.
%The best-fit parameters are $\alpha\sim 0.9$, $T_*\sim 5$ MeV, and $(\beta/H_*)^{-1}\sim 0.038$.
}\label{nocomred}
\end{figure*}

\subsection{On the SMBBH Interpretation for the Common Power-Law Excess}
\begin{figure*}[!htb]
\includegraphics[width=1\textwidth]{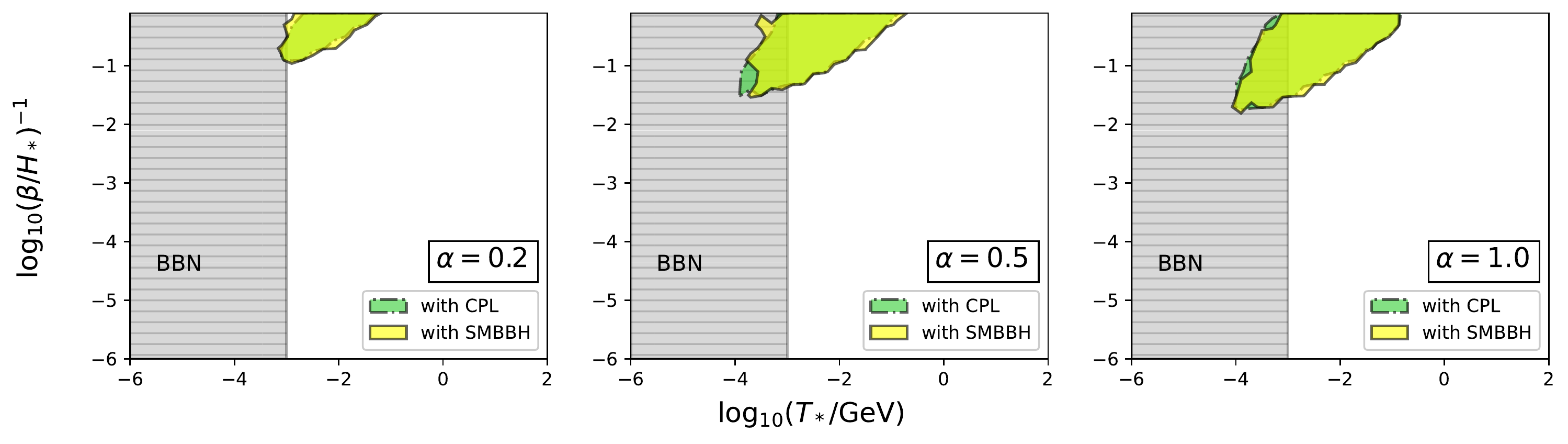}
\caption{{The PPTA exclusion contours (at 95\% C.L.) on FOPT parameters $\log_{10}(T_*/{\rm GeV})$ and $\log_{10}[(\beta/H_*)^{-1}]$ in the presence of a gravitational-wave background from supermassive binary black holes (SMBBH; in yellow) or a CPL process (green), for $\alpha=0.2$, 0.5, and 1.0.
%The best-fit parameters are $\alpha\sim 0.9$, $T_*\sim 5$ MeV, and $(\beta/H_*)^{-1}\sim 0.038$.
In both cases, we make use of the full-HD correlation in the analysis.}
}\label{SMBHB}
\end{figure*}

The constraints on the FOPT model parameters in the SMBBH + FOPT hypothesis are found to be similar with those of H3 and H4, as can be seen in Fig.~\ref{SMBHB}.
The power spectral density of the SMBBH signal is written as
\begin{align}
    S_{\rm SMBBH}(f) = \frac{A_{\rm SMBBH}^2}{12\pi^2} \left(\frac{f}{{\rm yr}^{-1}}\right)^{-\gamma_{\rm SMBBH}}{\rm yr}^{3},
\end{align}
where the prior for $\log_{10}(A_{\rm SMBBH})$ is uniformly distributed from $-18$ and $-14$ and $\gamma_{\rm SMBBH}$ is fixed at $13/3$.

\end{appendix}
\end{document}